\pdfoutput=1
\documentclass{ifacconf}

\usepackage{graphicx}      % include this line if your document contains figures
\usepackage{natbib}        % required for bibliography

\usepackage{amsmath,amssymb,amsfonts}
\usepackage{algorithmic}
\usepackage{xurl}
\usepackage{textcomp}
\usepackage{xcolor}

\newcommand{\barr}[2]{\begin{array}{#1}#2\end{array}}
\begin{document}
\begin{frontmatter}

\title{Energy-efficient Merging of Connected and Automated Vehicles using Control Barrier Functions\thanksref{footnoteinfo}} 

\thanks[footnoteinfo]{The information, data, or work presented herein was funded in part by the Advanced Research Projects Agency-Energy (ARPA-E), U.S. Department of Energy, under Award No. DE-AR0000837. The views and opinions of authors expressed herein do not necessarily state or reflect those of the United States Government or any agency thereof.}

\author[First]{Shreshta Rajakumar Deshpande} 
\author[First]{Mrdjan Jankovic} 

\address[First]{Southwest Research Institute, San Antonio, TX 78238, USA (e-mail: shreshta.rajakumardeshpande@swri.org;  mrdjan.jankovic@swri.org).}
%\address[Second]{Southwest Research Institute, Ann Arbor, MI 48180 USA (e-mail: mrdjan.jankovic@swri.org)}

\begin{abstract}                % Abstract of 50--100 words
Highway merges present difficulties for human drivers and automated vehicles due to incomplete situational awareness and a need for a structured (precedence, order) environment, respectively. In this paper, an unstructured merge algorithm is presented for connected and automated vehicles. There is neither precedence nor established passing order through the merge point. The algorithm relies on Control Barrier Functions for safety (collision avoidance) and for coordination that arises from exponential instability of stall-equilibria in the inter-agent space. A Monte Carlo simulation comparison to a first-in-first-out approach shows improvement in traffic flow and a significant energy efficiency benefit.
\end{abstract}


\end{frontmatter}
%===============================================================================
\section{Introduction}
Highway merges and interchanges are not easy to execute for human drivers. They are a source of significant traffic slowdowns as well as a high proportion of highway accidents. According to an estimate based on 2001 FARS (U.S. Fatality Analysis Reporting System) data, $18\%$ of interstate highway crashes and $11\%$ of fatalities occur at interchanges, even though these locations represent less than $5\%$ of total miles, see \cite{mccartt}. As a result, automating merges that employ vehicle-to-vehicle (V2V) and vehicle-to-infrastructure (V2I) communication have received significant research attention.
%, starting with the Intelligent Vehicle Highway System (IVHS) effort in the 1990s, see \cite{varaiya}. 
An extensive review of merge-related research up to 2014 presented in \cite{scarinci} divides algorithms into three categories: (a) fully automated, (b) cooperative adaptive cruise control (CACC) enabled, and (c) on-board display aid to human drivers. Often, some of the vehicles are considered to have formed platoons, see, for example, \cite{lu, wang2013}. 

Most merge algorithms, including the ones considered in this paper, are concerned with longitudinal trajectory planning only. It is assumed that a separate system can keep the vehicle in the lane. Even in the fully automated (centrally controlled) merge, a typical approach would be to consider vehicles in the merge lane requesters and the vehicles in the highway lane granters. The latter would create gaps or slots to allow a merge (\cite{lu, marinescu}). 
%This approach is also considered in the recent \cite{itu}, a United Nations specialized agency for information and communication technologies. 
With many vehicles in the merge and highway roads, the problem of finding and creating efficient, or at least reasonable, slots for everyone becomes combinatorial. To address this issue, Mixed Integer Programs (MIPs) are introduced to handle the discrete combinatorial component of the merge-related optimization. In \cite{mukai}, a Model Predictive Control (MPC) was combined with a MIP for the ego (merge road) vehicle. The MIP part selects an appropriate slot to merge -- the order is computed on-line -- while the existence of a sufficient gap is assumed. A MIP was also used in \cite{xu} in a planning-based strategy where a central entity controls all the vehicles. In this approach, the computational complexity is reduced by grouping vehicles into platoons, thus having a fewer number of options for the passing order that the MIP must consider.

In the case of equal priority of the main and merge lanes, methods such as ``logical ordering" (\cite{kanavalli}) or simply first-in-first-out (FIFO) (\cite{xiao2021, xiao2024}) can be used. When the vehicle order through the merge zone is fixed, the problem becomes convex and easier to solve (polynomial versus exponential scaling with the number of agents that could be either vehicles or platoons). This advantage of convexity will be featured in the approach described in this paper.

%%%%%%%%%%%%%%%%%%%%%%%%%%%%%%%%%%%%%%%%%%%%%%%%%%%%%%%%%%%%%%%%%%%%%%
\section{Proposed Contributions}

An algorithm for merge control is proposed in this paper: a centralized approach that does not establish explicit vehicle ordering through the merge zone. It uses control barrier functions (CBFs) (\cite{wieland, ames}) for collision avoidance while optimizing for all vehicles in the control zone (CZ) –- hence the name Centralized-CBF (C-CBF) Eco-merge algorithm. A pure decentralized algorithm would also provide safety, i.e. collision avoidance, but not absence of gridlocks as discussed in \cite{jankovicTCST, jankovicAR}. CBFs have been previously used in the merge control framework (\cite{xiao2021, xiao2024}), within an explicit (e.g. FIFO) order,  to prevent collisions, but not to determine the  passing order. In this paper, FIFO ordering combined with CBF-based collision avoidance is used as the benchmark to which C-CBF Eco-merge is compared. 

Like other merge control methods, C-CBF Eco-merge controller  requires V2V but not necessarily V2I communication. The controller is continuous with respect to the system state (vehicle positions, velocities), which facilitates distributed (replicated) computation because a small changes in the state would produce a small change in the control. In contrast, FIFO- or MIP-based algorithms could produce a detrimental change in control action with a minimal change in the state if it leads to a change in passing order. This is why, in general, explicit ordering requires a roadside coordinator to ensure all vehicles use the same order, even if the algorithm is otherwise decentralized.

The operation of the C-CBF Eco-merge algorithm requires minimal change to the standard V2V basic safety message (BSM) format -- see, J2735 \cite{SAE}. The BSM standard has each vehicle broadcast its position, velocity, acceleration, steering angle, size, etc., at $10$ Hz rate. In addition to BSM, our algorithm would require that each vehicle broadcasts its {\em current desired speed} obtained from its baseline longitudinal velocity controller, such as an adaptive cruise control (ACC) system.  

One advantage of the C-CBF Eco-merge algorithm is that vehicle size and mass can be included in the computation. Intuitively, human drivers do not expect a much larger vehicle (say, a semi truck) to change its speed much to accommodate a merge coordination, whether the semi is on the highway or the merging road. The algorithm allows the vehicle masses to impact the computed actions in a proportional manner. This is more difficult to accomplish with order-based algorithms, and in most cases size and mass differences are not considered for order setting.

Having mass and size information at its disposal, the C-CBF Eco-merge approach allows consideration of the energy impact. In particular, the coast-down (i.e., road load) information for light-duty vehicles published by the \cite{EPAtestdata} was leveraged to calculate the total energy impact of a merge scenario. Note that this paper just considers energy at the wheels and is powertrain-agnostic, thus excluding possible regenerative braking benefits of electrified vehicles. This paper shows that the unstructured C-CBF Eco-merge algorithm improves traffic flow and fuel efficiency, in several different metrics, over the FIFO approach. The fuel economy benefits is there even when all the vehicles are the same size/weight. That is, the benefit is not just due to heavier vehicles braking and accelerating less.

%%%%%%%%%%%%%%%%%%%%%%%%%%%%%%%%%%%%%%%%%%%%%%%%%%%%%%%%%%%%%%%%%%%%%%
\section{Problem Formulation}

\subsection{Problem Setting}
\begin{figure}[!b]
	\centering
	\includegraphics[width=\columnwidth]{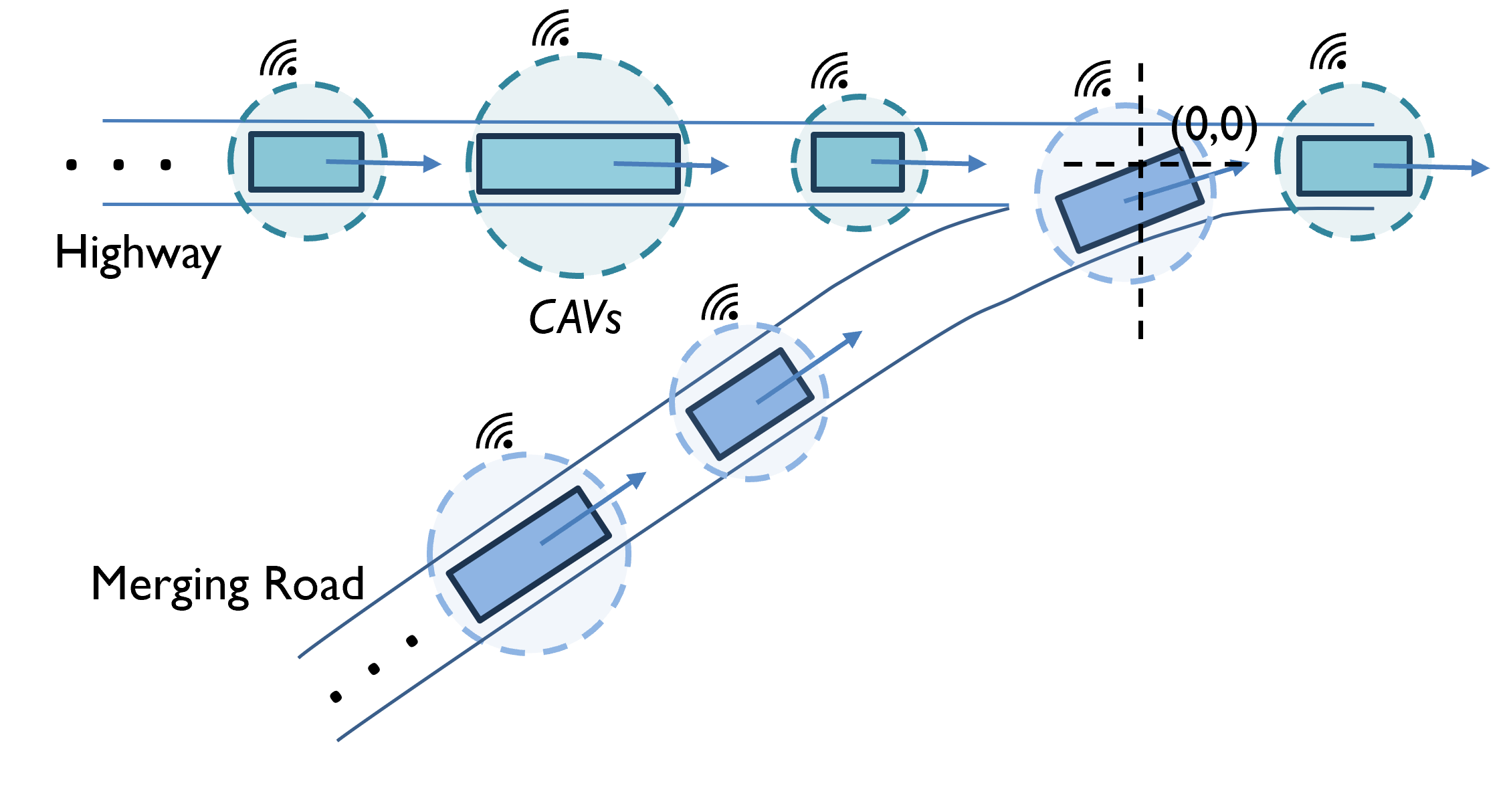}
	\caption{Short merge scenario with communicating CAVs.}
	\label{fig:short_merge_setup}
\end{figure}

The objective of the proposed Eco-merge algorithm is to improve the efficiency of merges for CAVs, both in terms of energy consumption and travel time, while assuring safety (collision avoidance). Fig. \ref{fig:short_merge_setup} illustrates the merge scenario considered. Here, the merge point is taken as the origin of the coordinate frame. As shown in Fig. \ref{fig:short_merge_setup}, the communicating CAVs that compose the traffic have different sizes (and masses) and may have different desired velocities. In the proposed scheme, each vehicle replicates the central controller computing the actions of all the CAVs within a virtual CZ. The controller requires knowledge about the size, locations, actual and desired velocity of each vehicle.

\subsection{Brief Review of Control Barrier Functions}
Control barrier functions have become a standard control design tool for systems with constraints. For a nonlinear system
\begin{equation} \dot x = f(x) + g(x) u \label{nls} \end{equation}
with the state $x \in R^n$, the control input $u \in {\mathcal{U}} \subseteq R^m$, and the admissible set ${\mathcal{C}} = \{x\in R^n: h(x) \ge 0\}$, the function $h(x)$ is a CBF if
\begin{equation} \max_{u \in {\mathcal{U}}} \left\{ \frac{\partial h}{\partial x}f(x) + \frac{\partial h}{\partial x}g(x) u + \lambda \cdot h(x) 
\right \} > 0  \label{CBF} \end{equation}
where $\lambda > 0$ is the barrier bandwidth, often available as a tuning parameter.  For computational efficiency, ${\mathcal{U}}$ is assumed to be a convex set. Consistently applying $u \in {\mathcal{U}}$ that satisfies the above inequality leads to  $h(x(t)) > h(x(0)) e^{-\lambda t}$, assuring invariance of the admissible set ${\mathcal{C}}$. 

In most cases, enforcement of the admissible set is done with a safety filter setup. This approach requires a performance (baseline) control input $u_0$ that achieves a desired objective (e.g., vehicle speed control in this paper). The safety filter is a quadratic program (or a convex program if the set ${\mathcal{U}}$ is not just simple upper/lower limits) that enforces the barrier constraint for the CBF $h(x)$ while staying close to the baseline control $u_0$:
\[ \barr{l}{ \min_{u \in {\mathcal{U}}}  \|u - u_0\|^2 \ \ {\rm such \ that} \\*[2mm]
 \frac{\partial h}{\partial x}f(x) + \frac{\partial h}{\partial x}g(x) u + \lambda \cdot h(x) \ge  0}
\]
This setup assures safety because, by definition, the quadratic program is feasible. Strict feasibility, due to the strict inequality sign in \eqref{CBF}, provides  Lipschitz continuity of the resulting control \cite{jankovicRCBF}. For more complex situations with multiple constraints, such as for the
multi-agent system discussed below, \cite{jankovicTCST} showed that proportional braking for all agents (vehicles, in this case) is a strictly feasible action.

In contrast to other merge control methods, in this paper the CBFs are used not only to assure collision avoidance, but also for vehicle coordination through the merge zone. The key features of the C-CBF safety filter are
\begin{itemize}
    \item Favorable (i.e. polynomial) computational scaling;
    \item Controller (Lipschitz) continuity;
    \item {Instability of stall-equilibria} in the inter-agent dynamics (see \cite{jankovicTCST, jankovicAR});
    \item Existence of a measure-0 stable manifold.
\end{itemize}

% In contrast to other merge-control methods, in this paper the CBFs are used not only to assure collision avoidance, but also for vehicle coordination through the merge zone. The key features of the C-CBF safety filter are: (1) Favorable (i.e. polynomial) computational scaling; (2) Controller (Lipschitz) continuity; (3) Instability of stall-equilibria in the inter-agent dynamics (see \cite{jankovicTCST, jankovicAR}), and (4) Existence of a measure-0 stable manifold.

In a simple intersection case that resembles the merge problem (\cite{jankovicTCST}), the speed of instability was shown to be proportional to the desired (RMS) speed of the vehicles and inversely proportional to their size. Instead of switching ahead/behind, a pair of vehicles traveling along the stable manifold are slowing down, trading off speed for predictability (\cite{jankovicAR}).  

\subsection{Modeling and Barrier Constraint Formulation}
In this work, the vehicles are modeled as single integrators, in contrast to a more standard double integrator approach with acceleration as the control input. Using velocity as the control input improves the rate of instability (i.e., merge speed), while vehicle accelerations are constrained inside the safety filter setup.

A vehicle $i$ is modeled as a disk of radius $r^i$. Its center motion is given by:
\begin{equation} \label{eq:state_eqn}
    \begin{aligned}
    \dot{s}_t^i=v_t^i=u_t^i, \quad \forall i=1,\dots,N_a
    \end{aligned}
\end{equation}
where $s_t^i$ is the distance of vehicle $i$ at time $t$ from the merge point,  $v_t^i$ are its velocity equal to its control action $u_t^i$, and $N_a$ is the number of vehicles in the CZ. The relative motion between CAVs $i$ and $j$ is given by:
\begin{equation} \label{eq:rel_motion_eqns}
    \begin{aligned}
    \dot{\xi}_t^{ij} &= v_t^{ij} \\
    \xi_t^{ij} &= \begin{bmatrix} x_t^i - x_t^j & y_t^i - y_t^j \end{bmatrix}^\mathsf{T} \\
    % \xi_t^{ij} &= \left[x_t^i - x_t^j, y_t^i - y_t^j \right]^\mathsf{T}
    v_t^{ij} &= \begin{bmatrix} v_t^{ix} - v_t^{jx} & v_t^{iy} - v_t^{jy} \end{bmatrix}^\mathsf{T}
    \end{aligned}
\end{equation}
where the X-Y location (resp. velocity) of a vehicle $i$, computed from the road geometry, is given by $\{x_t^{i},y_t^{i}\}$ (resp. $\{v_t^{ix},v_t^{iy}\}$); $\xi_t^{ij}$ is the center-to-center (vector) separation between two vehicles and $v_t^{ij}$ is their relative velocity. For collision avoidance, the inter-vehicle separation $||\xi_t^{ij}||$ must be greater than the sum of their respective barrier radii, scaled by a tunable safety margin $\beta$ (added for robustness). To achieve this, the barrier function $h_t(\xi_t^{ij})$ is defined:
\begin{equation} \label{eq:barrier_fn}
    \begin{aligned}
    h_t(\xi_t^{ij}) = (\xi_t^{ij})^\mathsf{T}\xi_t^{ij} - \left((1+\beta)\cdot \left(r^i + r^j \right) \right)^2
    \end{aligned}
\end{equation}

The CBF constraint is then given by:
\begin{equation} \label{eq:barrier_constraint}
    \begin{aligned}
    F_t^{ij} := \dot{h}_t(\xi_t^{ij}) &+ \lambda \cdot h_t(\xi_t^{ij}) \geq 0 \\
    A_t^{ij}u_t^{ij} &+ B_t^{ij} \geq 0
    \end{aligned}
\end{equation}
where $A_t^{ij} = 2\cdot (\xi_t^{ij})^\mathsf{T}$ and $B_t^{ij} = \lambda \cdot h_t(\xi_t^{ij})$. Each pair of vehicles generates one CBF constraint. Considering all the vehicles in the CZ, the total number of barrier constraints are $\frac{N_a(N_a-1)}{2}$.

%%%%%%%%%%%%%%%%%%%%%%%%%%%%%%%%%%%%%%%%%%%%%%%%%%%%%%%%%%%%%%%%%%%%%%
\section{Centralized-CBF Eco-merge Algorithm}

For safe and efficient merging, vehicles in the CZ should (1) avoid collisions with other vehicles and do so in a cooperative and comfortable manner, (2) travel close to their desired velocity (defined as a fixed set-point or a time-varying reference from a longitudinal velocity controller), and (3) account for heterogeneous traffic conditions (i.e., vehicles having different sizes or masses and operating in different traffic densities). The quadratic program (QP) defined in \eqref{eq:centralized_eco_merge_cost_fn_constr} has to be solved at discrete time instances. This fact is exploited to introduce a penalty on the rate of change of velocity -- the heavier the vehicle, the higher the cost. The C-CBF Eco-merge algorithm is formulated to simultaneously compute the velocity (control action) for all CAVs in the CZ:
\begin{equation} \label{eq:centralized_eco_merge_cost_fn_constr}
    \begin{aligned}
    \min_{u_k^1,\dots,u_k^{N_a}} &\sum _{i=1}^{N_a} \left\Vert u_k^i - v_{k0}^i\right\Vert^2 + m_{veh,sc}^i \cdot \left\Vert u_k^i - u_{k-1}^i\right\Vert^2 \\
    \text{subject to } &A_k^{ij}u_k^{ij} + B_k^{ij} \geq 0, \quad \forall i,j = 1,\dots, N_a, i\neq j
    \end{aligned}
\end{equation}
where $k$ denotes the discrete time instant corresponding to the current time $t=kT_s$ (where $T_s$ is the sampling time), $v_{k0}^i$ is the desired velocity of the vehicle $i$, and $m_{veh,sc} = \alpha \cdot m_{veh}$ denotes its scaled mass ($m_{veh}$ being the (actual) vehicle mass and $\alpha$ a scaling factor). Adding the rate-of-change terms to the cost does not impact feasibility or continuity, but  slows down somewhat the rate of instability. The matrices $\{A_k^{ij},B_k^{ij}\}$ are the terms from the CBF-based constraints defined in \eqref{eq:barrier_constraint}.   

The collision constraints can be augmented to include additional box constraints, including acceleration limits. In this work, the following acceleration limits are imposed: $a_k^i = \frac{u_k^i - u_{k-1}^i}{T_s} \in [-6, 5]$ m/s$^2$. The described optimization problem \eqref{eq:centralized_eco_merge_cost_fn_constr} was solved using qpOASES, a structure-exploiting active-set QP solver. The resulting policy is continuous and controls all the vehicles in the CZ without assigning them an explicit order. As noted earlier, a replica of this controller could run on-board each vehicle, avoiding the need for a roadside coordinator. The worst-case QP compute time per step for the 20-agent simulations is just $10$ ms on a laptop\footnote{$2.5$ GHz $13^{\text{th}}$ Gen Intel Core i7-13800H, $16$ GB RAM}, verifying its computational scalability.
\begin{rem}
    The arrival of a new CAV into the CZ may change the (implicit) merge order of the downstream traffic. This behavior was particularly observed for cases where there was large difference between the vehicles in terms of their sizes (masses) and speeds.
\end{rem}
%A distinctive feature of this formulation is the consideration of the mass (size) of the vehicles in the cost function of the optimization routine, which would enable its effective use in heterogeneous traffic -- containing light, medium and/or heavy-duty vehicles. Note that the algorithm would compute and apply lower accelerations for vehicles with larger $m_{veh}$. % The first term in the objective function penalizes deviation from the reference speed, while the second penalizes the rate of change of each vehicle's speed (effectively, acceleration).

% A noteworthy benefit from the Eco-merge algorithm developed is the expanded range of vehicle speeds and traffic densities at which automated merge maneuvers can be performed safely and comfortably. Simulations have shown that:
% \begin{itemize}
%     \item Traffic with high density ($1800$ veh/hr injected in each of the highway and merging roads),
%     \item CAVs having speeds up to: (1) $30$ m/s or $65$ mi/hr (with acceleration limits: $\{-6, 5\}$ m/s$^2$), (2) $20$ m/s or $45$ mi/hr (with acceleration limits: $\{-4, 3\}$ m/s$^2$)
% \end{itemize}
% can be handled by the proposed C-CBF approach without collisions (with $T_s = 0.05$ s, $\lambda = 0.25$, and the CZ starting at a reasonable $150$ m from the merge point).

Another feature of this controller is that it can be integrated with existing longitudinal velocity controllers (including Eco-driving controls such as \cite{deshpande2022real,wang2023connected,deshpande2024real}) as a safety filter that adds cooperative merging capability.

%%%%%%%%%%%%%%%%%%%%%%%%%%%%%%%%%%%%%%%%%%%%%%%%%%%%%%%%%%%%%%%%%%%%%%
\section{Evaluation and Results}

\subsection{Benchmark}
C-CBF Eco-merge was evaluated against a FIFO benchmark as the standard approach for merge and intersection management. In the FIFO approach, the first vehicle to enter the defined CZ is the first one out; this priority is assigned by a central coordinator. The FIFO approach is itself quite efficient and likely much better than human drivers in this respect.

Because FIFO is essentially non-cooperative (once the vehicle order is fixed), using velocity as the control variable produced significant  backward propagation of slowdowns (a.k.a. ``string instability"). Thus, for more realistic comparison, each FIFO vehicle was modeled as a double integrator with the acceleration as the control input. A second-order CBF constraint was used for collision avoidance. The constraints were formulated for higher priority  vehicles, while lower priority ones (those behind it) are ignored.  

In addition to the barrier constraints for safety, acceleration limits were also applied in the FIFO case (for fair comparison with the Eco-merge algorithm). Optionally, a slack variable can be used for feasibility guarantees.

\subsection{Evaluation Metrics}
In the context of merge simulations, travel time refers to the time taken by the last vehicle in the simulation to cross the merge point. Comparison between the C-CBF Eco-merge and the FIFO benchmark was performed using their travel (merging) times as well as the following powertrain-agnostic energy consumption metrics: Positive acceleration Kinetic Energy, Braking Energy and Total Energy Loss.

\subsubsection{Positive acceleration Kinetic Energy (PaKE)}
At each discrete time instant $k$, the PaKE of each vehicle (adapted from \cite{milkins1983comparison}) is defined as:
\begin{equation} \label{eq:PaKE_eqn}
    \begin{aligned}
    \frac{\sum _{k=0}^{N-1} m_{veh} \cdot \max \left(0, v_{k+1}^2 - v_k^2 \right)}{s_N}
    \end{aligned}
\end{equation}
where $s_N$ is the distance traveled by the vehicle.

\subsubsection{Braking Energy (BE)}
The BE of each vehicle is defined as:
\begin{equation} \label{eq:BE_eqn}
    \begin{aligned}
    \frac{\int_0^t \max \left(0, -m_{veh}\cdot a_t - {F_{loss,rl,t}} \right) \cdot v_t\ \mathrm{d}t}{s_N}
    \end{aligned}
\end{equation}
where $F_{loss,rl,t}$ is the road load loss at time $t$, given by: 
\begin{equation} \label{eq:F_loss_roadload}
    \begin{aligned}
    F_{loss,rl,t} = A_{dyno}+B_{dyno}\cdot v_t+C_{dyno}\cdot v_t^2
    \end{aligned}
\end{equation}
where $\{A_{dyno},B_{dyno},C_{dyno}\}$ are the vehicle-dependent dyno coefficients published by the \cite{EPAtestdata} for each vehicle they test. Note that the formulated BE metric only considers the braking action in excess of the coasting-based deceleration.

\subsubsection{Total Energy Loss (TEL)}
The TEL is defined as:
\begin{equation} \label{eq:nTEL_eqn}
    \begin{aligned}
    \frac{\int_0^t \max \left(F_{loss,brk,t}, F_{loss,rl,t} \right) \cdot v_t \ \mathrm{d}t} {s_N}
    \end{aligned}
\end{equation}
where $F_{loss,rl,t}$ is computed using \eqref{eq:F_loss_roadload} and $F_{loss,brk,t}$ denotes the loss from braking action, given by: 
\begin{equation*} \label{eq:F_loss_brk}
    \begin{aligned}
    F_{loss,brk,t} = -\min \left(0, a_t \right) \cdot m_{veh}
    \end{aligned}
\end{equation*}
\begin{rem}
    These definitions of the energy consumption metrics are at a vehicle-level and are normalized with respect to the respective distance traveled. The corresponding system-level metrics are computed by averaging them across all the $N_a$ vehicles in the CZ.
\end{rem}

\subsection{Monte Carlo Simulations for Heterogeneous Traffic} \label{sec:MC_sim_het_traffic}

\begin{table}[!b]
\centering
\caption{Monte Carlo simulation setup.}
\label{table:MC_sim_setup}
\begin{tabular}{cc}
\hline
\textbf{Parameter or Variable} & \textbf{Value} \\ \hline
No. of simulation runs & $500$ \\
Merge road angle & $30^\circ$  \\
Initial (desired) velocity & $v_{k0} \sim \mathcal{U}\left(20, 25\right)$ m/s  \\
Injection rate (per road) & $\sim \mathcal{U}\left(1100, 1200\right)$ veh/hr \\
Sampling time & $T_s = 0.1$ s \\
Vehicle mass & $m_{veh} \sim \mathcal{U}\left(m_{base}, 4 \cdot m_{base}\right)$ lbs \\
Vehicle (barrier) radius & $2-4$ m \\ % (linear scaling with $m_{veh}$)
% Barrier bandwidth & $\lambda = 0.25$ \\
Barrier margin & $\beta = 10\%$
\end{tabular}
\end{table}

\begin{figure}[!b]
	\centering
	\includegraphics[width=0.8\columnwidth]{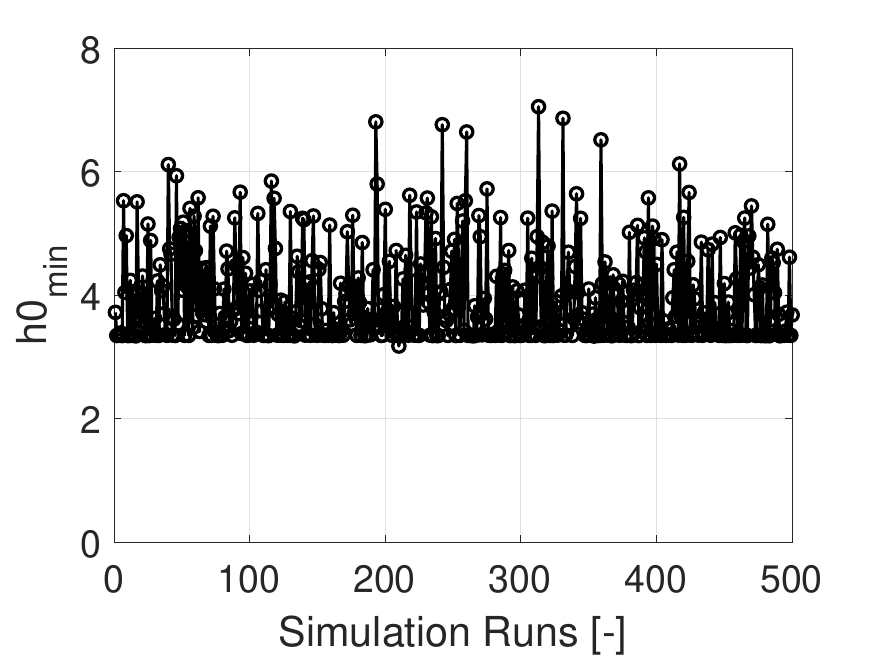}
	\caption{Minimum barrier distance for 500 20-vehicle Monte Carlo simulations.}
	\label{fig:MC_sim_evaluation_hmin}
\end{figure}

Monte Carlo simulations were performed to compare the Eco-merge algorithm against the FIFO benchmark. For this evaluation, $20$ vehicles were considered -- $10$ each on the highway and merge roads. The CZ includes $200$ m road sections before the merge point and $350$ m after it. The simulation setup is provided in Table \ref{table:MC_sim_setup}. The desired (initial) speeds for the CAVs are selected randomly (with uniform distribution) between $20-25$ m/s. The injection rate is also randomly selected (uniform) between $1100-1200$ vehicles per hour on each road. The chosen $m_{base}$ is $2375$ lbs and corresponds to a Mitsubishi Mirage; the largest vehicle mass corresponds to a Chevrolet Silverado EV which is 4 times heavier. The vehicle masses are selected randomly with uniform distribution between these two extremes. The target coast-down coefficients are obtained for these vehicles from the EPA. The sampling time is $T_s = 0.1$ s, consistent with BSM broadcasting rate. The vehicle (barrier) radius, having a range of $2-4$ m, is computed by linear scaling based on the selected $m_{veh}$. For fair evaluation, the same initial conditions and vehicle masses are used in the C-CBF and FIFO runs. The barrier bandwidth for the C-CBF is $\lambda = 0.25$ while FIFO used $\lambda_1 = 0.3, \lambda_2 = 2.0$.

Neither algorithm produced a collision across all the Monte Carlo simulation runs. For the C-CBF Eco-merge, this point is illustrated in Fig. \ref{fig:MC_sim_evaluation_hmin}. It shows the minimum inter-vehicle barrier distance $h0_{min}$ computed for all the vehicles during each simulation run ($h0_{min}$ corresponds to the barrier safety margin $\beta = 0$ in \eqref{eq:barrier_fn}, rather than $\beta = 0.1$ used in the  controller in simulations). 
%The lower bound of $h0_{min}$ is equal to the $\beta = 10\%$ radius margin assigned and is always greater than $0$ i.e., there are no barrier violations for all initial conditions simulated.

\begin{figure}[!t]
	\centering
	\includegraphics[width=\columnwidth]{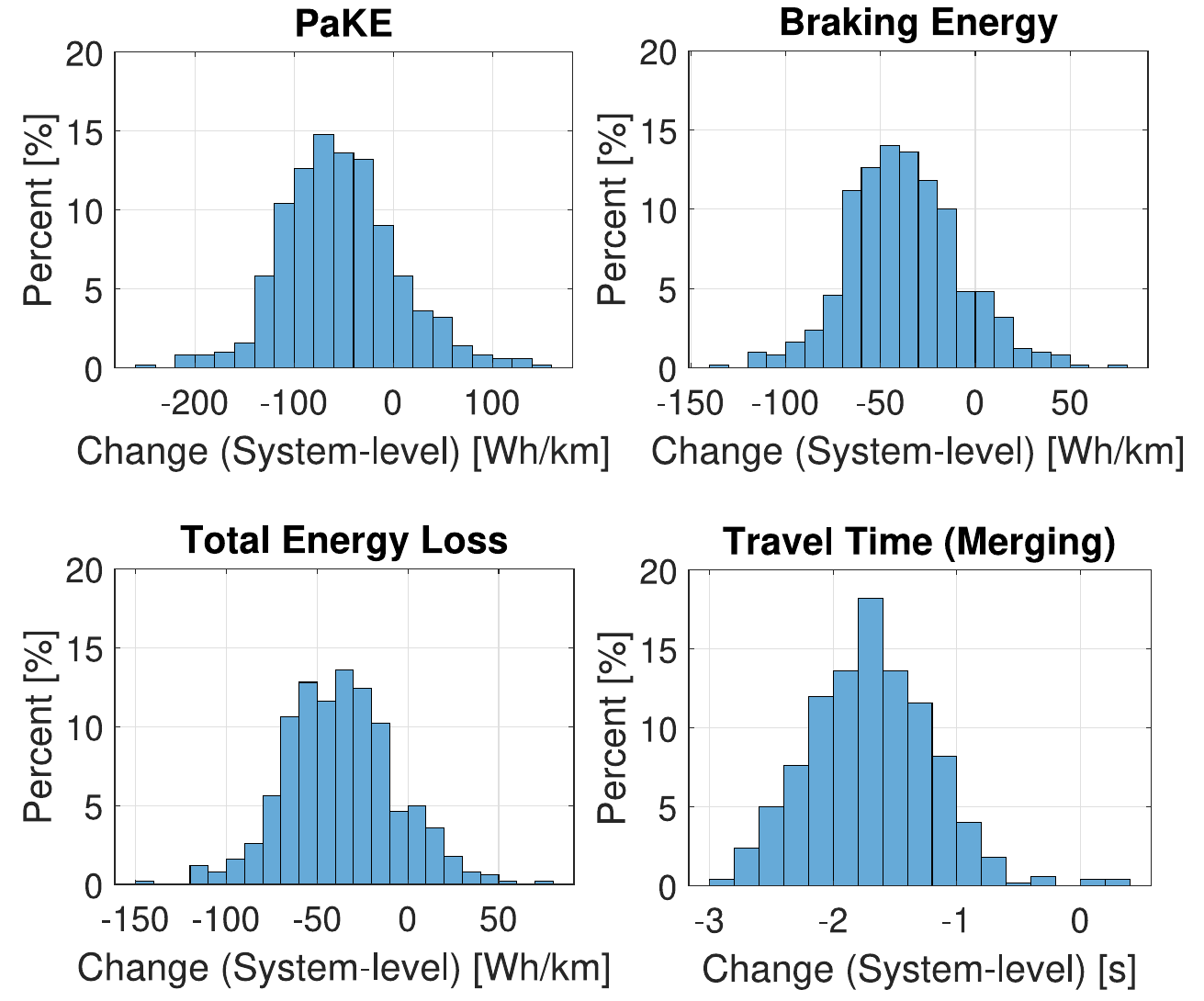}
	\caption{Monte Carlo simulation results -- histograms of key evaluation metrics.}
	\label{fig:MC_sim_evaluation_hist_plots}
\end{figure}

The key comparison metrics from the Monte Carlo simulations are shown in Tables \ref{table:MC_sim_eval_metrics_energy} and \ref{table:MC_sim_eval_metrics_flow}, and their histograms are visualized in Fig. \ref{fig:MC_sim_evaluation_hist_plots}. The C-CBF Eco-merge algorithm improves the system-wide performance across all the key evaluation metrics, compared to the FIFO benchmark. It reduces the PaKE (mean and median), a measure of energy used in (re)accelerations, by over $23\%$ and $28\%$ respectively. The BE (mean and median), a measure of energy losses from braking, is reduced by $34\%$ and $38\%$ respectively. The mean TEL, the metric that captures total energy losses, is reduced from $240$ Wh/km (FIFO) to $203$ Wh/km, i.e., by $15\%$ (median TEL is reduced by $16\%$). The simulation cases where the FIFO approach performed better than the C-CBF method are predominantly characterized by nearly symmetric initial conditions -- vehicles initialized close to the stable manifold discussed above.

Additionally, the C-CBF reduces system congestion as shown  in Table \ref{table:MC_sim_eval_metrics_flow}. It reduced the time  to merge by over $1.6$ s on average (nearly $4\%$). Further, the system-level average speed of all vehicles in the CZ increased by $5.8$\%, from $21$ m/s to $22.2$ m/s.

\begin{rem}
    The improvement in terms of travel time, in fact, increases the road losses experienced by the Eco-merge vehicles (traveling at higher speeds on average); when accounting for this phenomenon, the actual difference in TEL would be higher than indicated in Table \ref{table:MC_sim_eval_metrics_energy}.
\end{rem}

% \begin{table}[!t] % FIFO -- highway takes priority (fixed arrival rate)
% \centering
% \caption{Monte Carlo simulations (heterogeneous) -- key energy-related metrics.}
% \label{table:MC_sim_eval_metrics_energy}
% \begin{tabular}{ccc}
% \cline{2-3}
%  & \multicolumn{2}{c}{\textbf{Percent Change [\%]}} \\ \hline
% \textbf{Metric} & \textbf{Mean} & \textbf{Median} \\ \hline
% PaKE & $-25.3$  & $-27.5$  \\
% BE & $-14.5$ & $-17.7$ \\
% nTEL & $-4.7$ & $-6.0$
% \end{tabular}
% \end{table}
\begin{table}[!t]
\centering
\caption{Monte Carlo simulations (heterogeneous) -- key energy-related metrics.}
\label{table:MC_sim_eval_metrics_energy}
\begin{tabular}{ccc}
\cline{2-3}
 & \multicolumn{2}{c}{\textbf{Percent Change [\%]}} \\ \hline
\textbf{Metric} & \textbf{Mean} & \textbf{Median} \\ \hline
PaKE & $-23.0$  & $-26.2$  \\
BE & $-32.9$ & $-36.7$ \\
TEL & $-15.0$ & $-16.0$
\end{tabular}
\end{table}

\begin{table}[!t]
\centering
\caption{Monte Carlo simulation results (heterogeneous) -- key flow-related metrics.}
\label{table:MC_sim_eval_metrics_flow}
\begin{tabular}{cc}
\cline{2-2}
 & \textbf{Percent Change [\%]} \\ \hline
\textbf{Metric} & \textbf{Mean} \\ \hline
Travel Time & -4.0  \\
Average Velocity & +5.8 % 22.22 (C-CBF), 20.99 (FIFO)
\end{tabular}
\end{table}

\subsection{Analysis}
\begin{figure}[!t]
	\centering
	\includegraphics[width=\columnwidth]{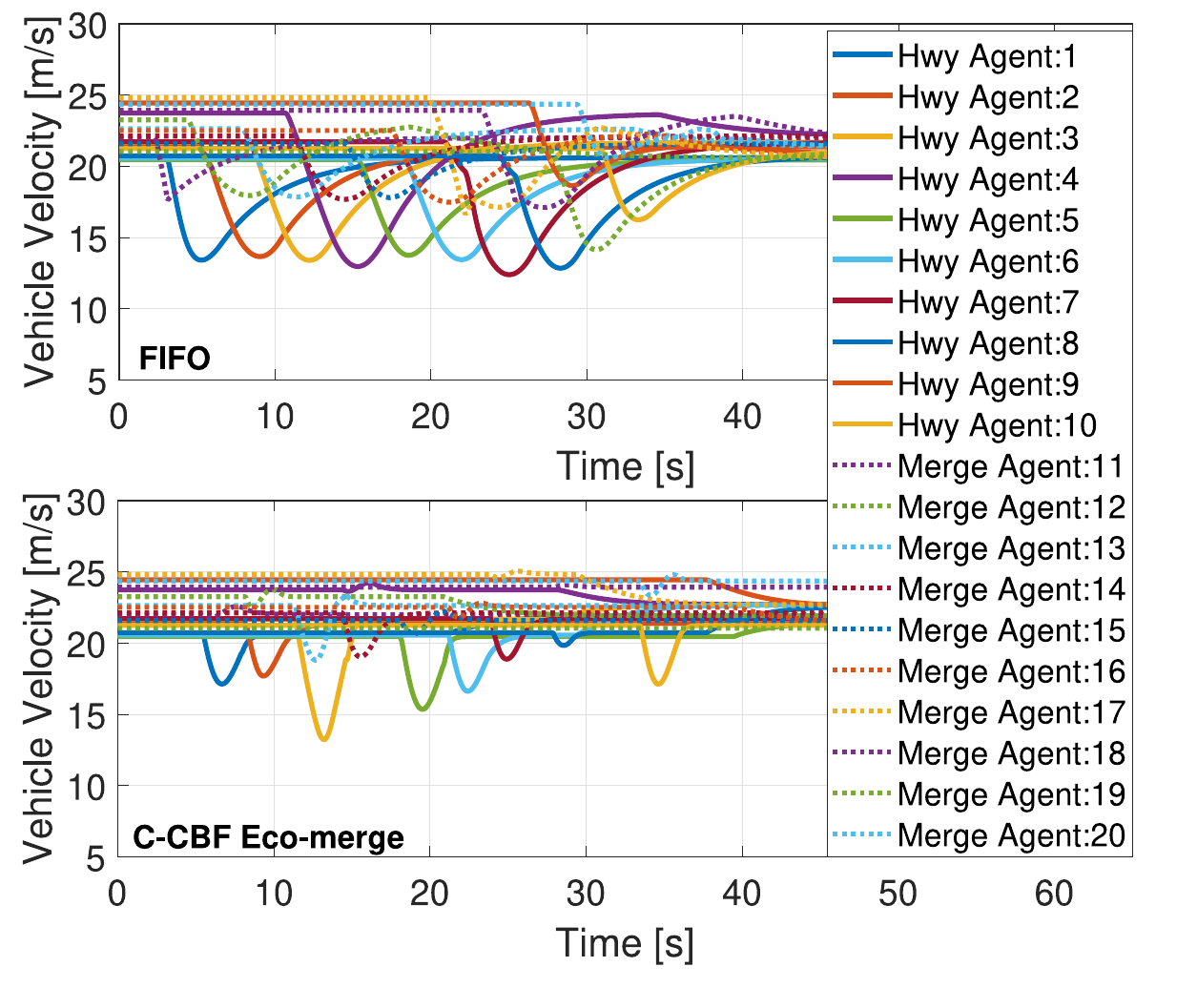}
	\caption{Comparison of 20-vehicle velocity traces: FIFO and C-CBF Eco-merge approaches.}
	\label{fig:FIFO_Ecomerge_vel_comp}
\end{figure}

Profiles of the vehicle velocities from one of the Monte Carlo runs, shown in Fig. \ref{fig:FIFO_Ecomerge_vel_comp}, are used to illustrate and analyze the difference between the velocity-based Eco-merge and the conventional FIFO merge methods. In the FIFO scheme, the order in which the vehicles merge is solely determined by their distance from the merge point at the time they enter the CZ -- only vehicles having (relatively) higher priority are considered while formulating collision constraints for a vehicle. As a result, FIFO methods provide orderly (zipper) merge sequences. On the other hand, in some instances a heavier vehicle traveling at a higher speed was forced to brake for a lighter, slower vehicle. This adversely affects the system-wide energy consumption. Additionally, such slowdowns propagate as the number of vehicles increases, thereby further increasing travel time.

% \begin{figure}[!b]
% 	\centering
% 	\includegraphics[width=0.8\columnwidth]{fifo_ecomerge_merge_order_comp_inj_rate_rv.pdf}
% 	\caption{Comparison of 20-vehicle merge order: FIFO and C-CBF Eco-merge approaches.}
% 	\label{fig:FIFO_Ecomerge_vel_comp}
% \end{figure}

In contrast to FIFO, the unordered control policy from the Eco-merge algorithm enables implicit coordination between all the CAVs in the network with a higher penalty on changing velocities of heavier vehicles. The result is smaller magnitude of velocity changes at a system-level. This is statistically corroborated by both the traffic flow- and energy-related metrics in Tables \ref{table:MC_sim_eval_metrics_energy} and \ref{table:MC_sim_eval_metrics_flow}. An interesting behavior exhibited by the vehicles of this method (not seen in FIFO) is that some lighter vehicles accelerate prior to the merge point to reduce the velocity drop (and consequent energy transactions) experienced by the following (heavier) vehicles.

\subsection{Monte Carlo Simulations for Homogeneous Traffic}

Monte Carlo simulations were then performed by considering homogeneous traffic i.e., all vehicles having the same size (mass). This helps in quantifying (and in a sense, separating) the system-level impact (energy and flow) of the C-CBF Eco-merge without traffic heterogeneity. Here, the vehicle mass was fixed  to $4500$ lbs (representative of a typical light-duty vehicle). Apart from this, the other random variables and simulation parameters remained the same as described in Section \ref{sec:MC_sim_het_traffic}. Tables \ref{table:MC_sim_fixed_eval_metrics_energy} and \ref{table:MC_sim_fixed_eval_metrics_flow} summarize the simulation results. Albeit a smaller improvement compared to that seen in Table \ref{table:MC_sim_eval_metrics_energy}, the total system-level energy consumption (as captured by the mean nTEL metric) is reduced from $217$ Wh/km (FIFO) to $187$ Wh/km. The slight decrease in the energy-related benefits compared to the heterogeneous case is expected -- attributable to the formulation of the C-CBF \eqref{eq:centralized_eco_merge_cost_fn_constr} and the insignificant impact of vehicle mass on the orderly FIFO benchmark. Notably, the C-CBF merge approach manages to reduce the system congestion similarly to the heterogeneous case.

\begin{table}[!t]
\centering
\caption{Monte Carlo simulations (homogeneous) -- key energy-related metrics.}
\label{table:MC_sim_fixed_eval_metrics_energy}
\begin{tabular}{ccc}
\cline{2-3}
 & \multicolumn{2}{c}{\textbf{Percent Change [\%]}} \\ \hline
\textbf{Metric} & \textbf{Mean} & \textbf{Median} \\ \hline
PaKE & $-22.1$  & $-25.3$  \\
BE & $-31.6$ & $-35.0$ \\
TEL & $-13.8$ & $-14.3$
\end{tabular}
\end{table}

\begin{table}[!t]
\centering
\caption{Monte Carlo simulations (homogeneous) -- key flow-related metrics.}
\label{table:MC_sim_fixed_eval_metrics_flow}
\begin{tabular}{cc}
\cline{2-2}
 & \textbf{Percent Change [\%]} \\ \hline
\textbf{Metric} & \textbf{Mean} \\ \hline
Travel Time & -3.9  \\
Average Velocity & +5.7 % 22.19 (C-CBF), 20.99 (FIFO)
\end{tabular}
\end{table}

%%%%%%%%%%%%%%%%%%%%%%%%%%%%%%%%%%%%%%%%%%%%%%%%%%%%%%%%%%%%%%%%%%%%%%
\section{Conclusions and Future Work}
This paper presents the C-CBF Eco-merge algorithm for highway merge coordination. It is fundamentally different from previous approaches because it is essentially unstructured -- there is no established precedence or passing order through the merge zone. The controller consists of only two components: a speed controller (e.g. ACC system), and a CBF safety filter that provides both collision avoidance and vehicle coordination. Benefits over a FIFO approach, in terms of traffic flow and energy efficiency, were presented. The energy efficiency comparison includes road load losses based on the EPA database. Future work would include improving robustness to ``unexpected behavior", a step that would allow human drivers in the mix.

\bibliography{ifacconf}             % bib file to produce the bibliography

%%%%%%%%%%%%%%%%%%%%%%%%%%%%%%%%%%%%%%%%%%%%%%%%%%%%%%%%%%%%%%%%%%%%%%
% \appendix
% \section{A summary of Latin grammar}    % Each appendix must have a short title.

\end{document}